\documentclass[12pt]{article}
\usepackage[DIV13]{typearea}

\usepackage{graphicx}
\usepackage{latexsym}
\usepackage{epsf}
\usepackage{epsfig}
\usepackage{subfigure}
\usepackage{amssymb,amsmath}
\usepackage[english]{babel}
\usepackage{amsfonts}
\usepackage{amssymb}
\usepackage{graphics}
\usepackage{mathrsfs}
\usepackage{verbatim}
\usepackage{float}
\usepackage{slashed}
\usepackage{comment}

\title{
\vspace*{5mm} \Large\textbf{Minimum Length - Maximum Velocity}
\vspace*{1.0cm}
\author{\textbf{Boris Panes\footnote{boris.panes@desy.de}}\\\\
\normalsize\emph{II. Institute for Theoretical Physics, University of Hamburg.}\\
\date{\today}}}

\begin{document}

\setcounter{page}{0}
\maketitle

\begin{abstract}
We study a framework where the hypothesis of a minimum length in space-time is complemented with the notion of reference frame invariance. It turns out natural to interpret the action of the obtained reference frame transformations in the context of doubly special relativity. As a consequence of this formalism we find interesting connections between the minimum length properties and the modified velocity-energy relation for ultra-relativistic particles. For example we can predict the ratio between the minimum lengths in space and time using the results from OPERA about superluminal neutrinos.
\end{abstract}

\thispagestyle{empty}

\newpage

\setcounter{page}{1}

\section{Introduction}

In the literature we can find a variety of theoretical arguments that support the existence of a fundamental minimum length (ML) scale, see~\cite{Garay:1994en} and references therein for a Quantum-Gravity review. Such a ML could be universal, like the fundamental length of strings in String Theory~\cite{LBL-28557,IFUP-TH-46-89} or the length scale arising from the discretization of the area in Loop Quantum Gravity~\cite{hep-th/9203079,PRINT-92-0447 (PITTSBURGH)}. But we could also consider non-universal ML scales, for example associated to the lower limit of the position uncertainty of fundamental particles~\cite{hep-th/9301067}, which could be given by the Compton radius $R_{C}=\hbar/M$ or the Schwarzschild radius $R_{S}=2GM$ depending on whether $M < M_{Pl}$ or $M > M_{Pl}$ respectively.  

Considering the potential insights into fundamental theories that the observation of such ML effects could give us, several bottom-up approaches have been proposed in order to incorporate the ML hypothesis on the basis of standard physics. Modifications of the uncertainty principle~\cite{hep-th/9301067,Kempf:1996nk}, non-commutative geometry~\cite{Piacitelli:2010bn}, extensions of Lorentz transformations~\cite{AmelinoCamelia:2000mn}, or several of the previous ideas applied in the context of QFT~\cite{Mimasu:2011sa,Kober:2010sj,Hossenfelder:2007re,Hossenfelder:2006cw}, are probably the most studied approaches. From this vast family of ML models we consider two particular approaches:

\begin{itemize}
 \item \textbf{The Generalized Uncertainty Principle (GUP)}: in this approach the position-momentum uncertainty principle is modified in such a way that the dispersion of the position operator is bounded from below. In this way we can interpret the ML as the smallest accuracy that we can reach by measuring the localization of a state. The consistency of this approach is supported by the quasi-position states formalism developed in~\cite{Kempf:1996nk} and~\cite{Kempf:1994su}. In this formalism any physical state is expanded in the basis of quasi-position wave functions, which have by definition a finite width. 
 \item \textbf{Doubly Special Relativity (DSR)}: in this approach the standard Lorentz transformations are modified in order to incorporate a fundamental length scale. These transformations should be consistent with the invariance of the measurement of this length. The explicit DSR transformations are not uniquely defined, however we can find in~\cite{AmelinoCamelia:2000mn} and~\cite{AmelinoCamelia:2000ge} the minimal criteria that these transformations should satisfy. In this context the particular methods both to define and to measure the ML are only restricted by the requirement of observer independence. 
\end{itemize}

We focus the attention on these two approaches because they are complementary from a quantum-relativistic point of view. Also we can find several references~\cite{Hossenfelder:2005ed,Magueijo:2002am,Cortes:2004qn} that put in evidence the correlation between the mathematical structures of each approach. Considering this natural interplay between the GUP and DSR we study a common framework that intends to connect both ideas. This allows us to investigate the correlations between the quantum-relativistic observables associated with a ML hypothesis.

The paper is organized as follows. In Section 2 we study the GUP formalism in four dimensions in order to determine the conditions for the existence or non-existence of finite MLs both in space and in time. In Section 3 we analyze the unitary transformations of the position and momentum operators that are compatible with the invariance of the GUP formalism. These transformations are naturally interpreted in the context of DSR\@. In this sense we propose that the fundamental length that is invariant under DSR is a quantum limit determined from the GUP\@. In Section 4 we investigate the classical behavior of the four-position and four-momentum vectors under the action of the DSR transformations obtained in Section 3. As a consequence of these transformations we are able to obtain a modified expression for the velocity in terms of energy and momentum, that does not rely on the modification of the special relativity (SR) dispersion relation. Considering this expression for the velocity, in Section 5 we propose a solution for two problems related to the DSR transformations. In Section 6 we study the phenomenology of this formalism, which allows us to relate the value of the MLs in space and time with the maximum velocity of ultra-relativistic particles. We reserve the Section 7 for conclusions.

\section{Minimum Length in space-time}

In order to make contact with several studies on ML using the GUP approach~\cite{Kempf:1996nk,Kempf:1994su,Hossenfelder:2004gj,Hossenfelder:2003jz,Ali:2010yn,Ali:2011ap}, we consider as starting point the modification of the commutation rule between the operators of position $x^{\mu}$ and momentum $p^{\mu}$. We parameterize these modifications assuming the existence of auxiliary functions $\rho^{\mu}(p)$ obeying canonical commutation rules with the position operators (throughout this work we set $\hbar=1$ and $c=1$), 

\begin{equation}
  [x^{\mu}, \rho^{\nu}(p)] = -i\eta^{\mu\nu} \hspace{2mm} \Rightarrow \hspace{2mm} [x^{\mu}, p^{\nu}] = -i \eta^{\mu\alpha}\frac{\partial p^{\nu}}{\partial \rho^{\alpha}}.
\label{MC4D}
\end{equation}

A sufficiently general parameterization of the auxiliary functions $\rho^{\mu}(p)$ that on the one hand will simplify future computations, but on the other hand will also allow us to identify clearly the realm of our approach, is given by 

\begin{equation}
  \rho^{\mu}(p) = (F(p^{0},T), (p^{i}/p)F(p,L)) \hspace{4mm} \text{with} \hspace{2mm} p \equiv |\vec{p}| \hspace{2mm} \text{and} \hspace{2mm} i=1..3, 
\label{RHOP}
\end{equation}

\noindent where $T$ and $L$ are parameters with units of length, introduced in order to construct dimensionless combinations with the energy and momentum variables. In principle we could use different functions $F(p^{0},T)$ and $G(p,L)$ in Eq.~(\ref{RHOP}), but we consider the simplest case where both functions are the same. The function $F$ is assumed to be differentiable and invertible. We also require that in the limit $p^{\mu} \rightarrow 0$ the standard commutation rules are recovered, i.e., $\rho^{\mu}(p)\rightarrow p^{\mu}$. Aside from these requirements, both the parameters $T$, $L$, and the function $F$ are degrees of freedom of this formalism.  

Given the parameterization~(\ref{RHOP}) and assuming that $[p^{\mu},p^{\nu}]=0$ it can be shown (following  Eqs. (27) to (32) of Ref.~\cite{Kempf:1996nk}) that spatial position operators satisfy the standard commutation rule $[x^{i},x^{j}]=0$. Furthermore, we can see that $[x^0,p^{i}]=0$ and $[x^{i},p^{0}]=0$, which allow us to find that $[x^0,x^{i}]=0$. Thus, the parameterization~(\ref{RHOP}) together with $[p^{\mu},p^{\nu}]=0$ implies a four dimensional commutative space-time. Considering these properties plus the assumption of rotational symmetry, we find that the GUPs for time and spatial components are

\begin{eqnarray}
 \Delta x^{0} \Delta p^{0} &\geq& \frac{1}{2}|\langle \frac{1}{F'(p^{0})}\rangle| \hspace{21mm} \text{with} \hspace{4mm} F'(p^{0}) = \frac{\partial F(p^{0},T)}{\partial p^{0}}\nonumber \\
 \Delta x^{i} \Delta p^{i} &\geq& \frac{1}{2}|\langle \frac{1}{3F'(p)} + \frac{2p}{3F(p)}\rangle| \hspace{4mm} \text{with} \hspace{4mm} F'(p) = \frac{\partial F(p,L)}{\partial p}.
\label{GUPS}
\end{eqnarray}

In the case that wave functions are rotationally invariant, we find non-trivial relations only for $\Delta x^{i} \Delta p^{i}$, but not for $\Delta x^{i} \Delta p^{j}$ with $i \neq j$. The particular form of the $\rho^{\mu}(p)$ map given in~(\ref{RHOP}) is fundamental to find this result, but for complete consistency we should also incorporate the rotational symmetry explicitly into dynamical equations. Considering only the non-trivial GUPs we are able to investigate the possible existence of minimum lengths both in space and in time, and we will define as a minimum length scenario (MLS) the symmetric situation when both scales have a finite value.

The structure of the GUPs given in Eq.~(\ref{GUPS}) indicates that the computation of MLs in space and in time can be developed independently of each other. This decoupling between space and time components allows us to study the ML in space coordinates following the formalism of~\cite{Kempf:1996nk}. Furthermore, the corresponding computation of the ML in the time coordinate can be done analogously. As a remarkable feature of this formalism we are able to work with functions $F(p)$ defined at every order in $p$. Below, we explain only the general features of this procedure, considering space and time coordinates. For more details we refer the reader to reference~\cite{Kempf:1996nk}. 

We start by considering the limiting case of GUPs given in Eq.~(\ref{GUPS}), when the product between the uncertainties in position and momentum is minimized. These GUP equations can be translated into equations for wave functions, which are denominated ``squeezed equations''\@. For the space-time scenario that we are considering here, these equations are 

\begin{eqnarray}   
      (i\frac{\partial}{\partial \rho} + i k_{L}p(\rho))\psi_{k_{L}}(\rho)  &=& 0 \hspace{2mm} \text{with} \hspace{2mm} k_{L} = \frac{1}{3}\sum_{i}\frac{\Delta x^{i} \Delta p^{i}}{(\Delta p^{i})^{2}} \nonumber \\
      (i\frac{\partial}{\partial \rho^{0}} + i k_{T}p^{0}(\rho^{0}))\psi_{k_{T}}(\rho^{0}) &=& 0 \hspace{2mm} \text{with} \hspace{2mm} k_{T} = \frac{\Delta x^{0}\Delta p^{0}}{(\Delta p^{0})^{2}},  
\label{squeezed}
\end{eqnarray}

\noindent where we have reduced the equations in the three dimensional space to only one equation in terms of the norm of the vector $\vec{\rho}$. We write the equations in the $\rho$ space for simplicity, since the action of the $x$ operator is given by a simple partial derivative. In the definition of $k_{L}$ the three terms of the sum are equal, given the rotational symmetry. The quantities $k_{T}$ and $k_{L}$ parameterize the possible solutions of the squeezed equations. The wave functions $\psi_{k_{L}}(\rho)$ and $\psi_{k_{T}}(\rho^{0})$ represent the states that live on the boundary of the region allowed by the GUPs. In order to proceed with the next steps we assume that the definition of the function $F$ allows analytic solutions of the above equations. 

Considering the explicit form of the normalized solutions of Eqs.~(\ref{squeezed}) we are able to evaluate the dispersion of space-time operators, $\Delta x(k_{L})$ with $\Delta x= (\Delta \vec{x}^{2})^{\frac{1}{2}}$ and $\Delta x^{0}(k_{T})$. Since we are interested in the lowest values of these dispersions we have to minimize with respect to the $k$ parameters. This minimization yields the values of the MLs in space and time.

Applying this procedure for four different choices of $F$, two bounded and two unbounded, we have noticed an important issue. We obtain finite values only in the bounded cases. The unbounded functions yield to vanishing minimum values. Two representative functions for each behavior are given in Table~\ref{maps}. One of the unbounded representatives is chosen to be the most slowly increasing unbounded function, the logarithm, just to emphasize our point. Although this behavior looks very general and not limited to our examples, we have not been able to find a proof. Despite of the lack of a general proof, we conjecture that the bounded property of the function $F$ is fundamental in order to obtain a minimum length scenario.  

\begin{table}[ht]
        \centering
        \begin{tabular}{c|c}
            Bounded maps & Unbounded maps \\
             \hline \\
              $F(p,L) = \frac{1}{L^2 p}\big{(} \sqrt{1 + 2(Lp)^2} -1 \big{)}$ & $F(p,L) = \frac{1}{L}\ln (1+Lp)$ 
                    \\                  
                  & \\
              $F(p,L) = \frac{1}{L}\tanh(Lp)$ & $F(p,L) = \frac{1}{L}{\rm arsinh} \left(Lp \right)$ 
                    \\ 
                  & \\
             \hline \\
        $\Delta x_{min} \sim L$, $\Delta x^{0}_{min} \sim T$ & $\Delta x_{min}=0$, $\Delta x^{0}_{min} = 0$       
        \end{tabular}
\caption{Representative functions for the GUP formalism. In the bounded cases we use the examples from~\cite{Kempf:1996nk} and~\cite{Hossenfelder:2003jz}. For the unbounded regime there are no examples in the GUP literature. The functions $F(E,T)$ are defined analogously. The factor of proportionality between $\Delta x_{min}$ and the parameter $L$ is of order one, and the same is applicable for the relation between $\Delta x^{0}_{min}$ and $T$.}
\label{maps}
\end{table}

It is interesting to notice that for a bounded interval of $\rho(p)$'s, there do exist formal position eigenstates that form a set of all eigenbases to the self-adjoint extensions of the position operator~\cite{Kempf:1996nk,Kempf:1994su}. The eigenstates corresponding to the same self-adjoint extension are normalizable and have a discrete spectrum, where the spacing between eigenvalues is given by integer multiples of $2\Delta x_{min}$. In principle, we could be tempted to interpret this result as a discretization of position space. However, these eigenstates do not correspond to physical states. For instance, they have an infinite momentum and furthermore they are not in the domain of the modified Heisenberg algebra. Thus, instead of using position eigenstates to expand physical wave functions we must use quasi-position states, which have a minimum uncertainty in position that respects the minimum length limit of the modified Heisenberg algebra.

\section{Reference frame invariance of the GUP formalism}

The basis of the GUP formalism is given by the commutator structure of Eq.~(\ref{MC4D}). In this section we would like to find a set of unitary transformations that keeps this structure invariant. In order to interpret these transformations as reference frame transformations we consider as a guide the structure of Lorentz transformations.  

Let us consider the unitary transformation $U(\omega)= e^{-\frac{i}{2}\omega_{\mu\nu}J^{\mu\nu}}$, with $J^{\mu\nu}$ self-adjoint operators that satisfy the standard Lorentz algebra. Under the infinitesimal action of $U(\omega)$ we can write the expression for the transformation of an operator $z^{\mu}$ in the standard way,

\begin{equation}
 z'^{\mu} = z^{\mu} - \frac{i}{2}\omega_{\alpha\beta}[J^{\alpha\beta},z^{\mu}] + O(\omega^2),  
\end{equation} 

\noindent where $z^{\mu}$ represents the operators $x^{\mu}$, $p^{\mu}$ or $\rho^{\mu}(p)$. We notice that the particular transformation of each operator is governed by the commutation rule between this operator and the generator of the unitary transformations $J^{\mu\nu}$. Thus, in principle we have some freedom to choose how the operators transform. In this work this possibility is explicitly considered because it turns out that the standard Lorentz transformations are not consistent at all with some of the operator properties derived from the GUP formalism. 

Indeed, from the previous chapter we learnt that the operator $\rho^{\mu}(p)$ has to be bounded in order to obtain a MLS\@. This requirement is not compatible with standard Lorentz transformations. Moreover, considering this restriction and Eq.~(\ref{MC4D}) we notice that standard transformations of the $x^{\nu}$ operators are incompatible with reference frame invariance. Hence, the operator that is not directly restricted is the momentum operator $p^{\mu}$, whose transformation can be taken as the standard one without explicit contradiction to the commutator structure of the GUP formalism or the reference frame invariance. Thus, by definition we consider $p^{\mu}$ as an operator that transforms like a standard Lorentz vector, with the corresponding commutator rule with the generator $J^{\mu\nu}$,

\begin{equation}     
  [J^{\mu\nu},p^{\alpha}] = i(\eta^{\mu\alpha}p^{\nu} - \eta^{\nu\alpha}p^{\mu}).
\label{jotap}
\end{equation}

Once we have fixed the action of the Lorentz generator on the momentum space, the corresponding action on the $\rho^{\mu}(p) $ operator can be derived immediately,

\begin{eqnarray}     
  [J^{\mu\nu},\rho^{\alpha}] &=& [J^{\mu\nu},p^{\beta}]\frac{\partial \rho^{\alpha}}{\partial p^{\beta}} \nonumber \\
                             &=& i(\eta^{\mu\beta}p^{\nu} - \eta^{\nu\beta}p^{\mu})\frac{\partial \rho^{\alpha}}{\partial p^{\beta}} \nonumber \\
  &=& i(p^{\nu}\frac{\partial \rho^{\alpha}}{\partial p_{\mu}} - p^{\mu}\frac{\partial \rho^{\alpha}}{\partial p_{\nu}}).
\label{jotarho}
\end{eqnarray}

In order to obtain a consistent commutator rule between $J^{\mu\nu}$ and $x^{\alpha}$ we consider the Jacobi identity that relates these three operators,

\begin{equation}
  [x^{\alpha},[J^{\mu\nu},\rho^{\beta}]] =  [[x^{\alpha},J^{\mu\nu}],\rho^{\beta}]+[J^{\mu\nu},[x^{\alpha},\rho^{\beta}]].
\end{equation}

Assuming the validity of Eq.~(\ref{MC4D}) and considering the expression for the commutator between $J^{\mu\nu}$ and $\rho^{\beta}$, derived from Eq.~(\ref{jotap}), we derive the condition 

\begin{equation}
  [[x^{\alpha},J^{\mu\nu}],\rho^{\beta}] = \eta^{\alpha\omega}\frac{\partial}{\partial \rho^{\omega}}\big{(}p^{\nu}\frac{\partial \rho^{\beta}}{\partial p_{\mu}} - p^{\mu}\frac{\partial \rho^{\beta}}{\partial p_{\nu}} \big{)},
\end{equation}

\noindent where we have used that $[J^{\alpha\beta}, \eta^{\mu\nu}] = 0$. A solution to this condition can be found if we assume that the commutator between $J^{\mu\nu}$ and $x^{\alpha}$ depends on $x^{\mu}$, $p^{\mu}$ and $\rho^{\mu}(p)$, which is clearly an extension of the standard scenario where this commutator depends only on $x^{\mu}$. Considering this assumption we find that a solution for this commutator is given by

\begin{equation}
 [J^{\mu\nu},x^{\alpha}] = i x^{\omega}\frac{\partial}{\partial \rho_{\alpha}}\big{(}p^{\mu}\frac{\partial \rho_{\omega}}{\partial p_{\nu}} - p^{\nu}\frac{\partial \rho_{\omega}}{\partial p_{\mu}}\big{)}.
\label{jotax}
\end{equation}

It is possible to show that using the commutation rules~(\ref{jotap}),~(\ref{jotarho}) and~(\ref{jotax}) the GUP structure is invariant under the unitary transformation $U(\omega)$. For example we obtain that

\begin{eqnarray}
  [x'^{\mu},\rho'^{\nu}] &=& [x^{\mu} - \frac{i}{2}\omega_{\alpha\beta}[J^{\alpha\beta},x^{\mu}] + O(\omega^2),\rho^{\nu} - \frac{i}{2}\omega_{\alpha\beta}[J^{\alpha\beta},\rho^{\nu}] + O(\omega^2)] \nonumber \\
  &=& [x^{\mu},\rho^{\nu}] -\frac{i}{2}\omega_{\alpha\beta}\big{(}[x^{\mu},[J^{\alpha\beta},\rho^{\nu}]] + [[J^{\alpha\beta},x^{\mu}], \rho^{\nu}]\big{)} + O(\omega^2) \nonumber \\
  &=& [x^{\mu},\rho^{\nu}] + O(\omega^2).
\end{eqnarray}

From this computation we can notice that the invariance of the commutator between $x^{\alpha}$ and $\rho^{\beta}$ follows from the Jacobi identity. Similar cancellations have to occur order by order in $\omega$, obeying the same kind of consistency relations.

Given the commutator~(\ref{jotax}), in principle we would be able to find the transformation of the operator $x^{\mu}$ for a finite $\omega$. However, as the commutator is non-linear on $p^{\mu}$ the analytic expression for the finite transformation is not easy to find. In order to circumvent this practical inconvenience we search for a combination between $x^{\mu}$ and $p^{\nu}$ satisfying a simpler commutation rule with $J^{\mu\nu}$. Consider the equation

\begin{equation}
 [J^{\mu\nu},x^{\alpha} f_{\alpha}{^{\beta}}(p)] = i(\eta^{\mu\beta}x^{\alpha} f_{\alpha}{^{\nu}}(p) - \eta^{\nu\beta}x^{\alpha} f_{\alpha}{^{\mu}}(p)),
\end{equation}

\noindent i.e., we require that the combination $x^{\alpha}f_{\alpha}{^{\beta}}(p)$, with $f_{\alpha}{^{\beta}}(p)$ unknown, satisfies a commutation rule with $J^{\mu\nu}$ analogous to the standard case. Using the commutation rules~(\ref{jotap}) and~(\ref{jotax}) we find that a solution to this equation is given by

\begin{equation}
  f_{\mu}{^{\nu}}(p) = \frac{\partial \rho_\mu(p)}{\partial p_\nu}.  
\end{equation}
 
With the help of this object we can find the transformation of the operators $p^{\mu}$ and $x^{\mu}$ for a finite $\omega$,

\begin{eqnarray}
     p'^{\mu} &=& \Lambda^{\mu}{_{\nu}}(\omega) p^{\nu} \nonumber \\
     x'^{\alpha}f_{\alpha}{^{\mu}}(p') &=& \Lambda^{\mu}{_{\nu}}(\omega) x^{\alpha}f_{\alpha}{^{\nu}}(p),      
\label{dsrtra}
\end{eqnarray}

\noindent where $\Lambda^{\mu}{_{\nu}}(\omega)$ are the standard matrices that represent the Lorentz transformations for a given set of parameters $\omega_{\mu\nu}$. The interpretation of these transformations and their parameters in terms of physical quantities is reserved for the next section. For the moment it becomes natural to extend these transformation rules given in terms of operators directly to the classical variables associated with the momentum and position. These transformations could be associated with the general subject of DSR~\cite{AmelinoCamelia:2010pd} because they arise from the requirement that the quantum MLs given by the GUPs are reference frame invariant. To our knowledge this particular approach to derive the transformations starting from the GUP structure is novel. In particular it turns out that the invariant dispersion relation for the four-momentum vector is the same as in SR ($E^{2} = p^2 + m^2$), which is not the case in most DSR studies~\cite{Magueijo:2002am,AmelinoCamelia:2010pd,Magueijo:2003gj,Judes:2002bw}. In this GUP/DSR approach we find that the main modifications are produced in the  position transformations, although as we will see these modifications will propagate to the relation between energy, momentum and velocity. Similar transformations for the position space can also be found in~\cite{Hossenfelder:2006rr} and~\cite{Kimberly:2003hp}. 

\section{Doubly Special Relativity transformations}

In order to analyze the effects of the transformations~(\ref{dsrtra}) at the classical level, we consider the action of active transformations on the four-momentum and four-position vectors. Consider a massive particle at rest, located at the origin of some reference frame. We define its four-momentum and four-position vectors in the standard way,

\begin{equation}
  q^{\mu} = (m,0,0,0)  \hspace{2mm} \text{and} \hspace{2mm} y^{\mu} = (t,0,0,0). 
\label{atrest}
\end{equation}

Defining a DSR boost in analogy with Lorentz boosts we consider the transformation obtained by setting $\omega_{0i} = \beta_{i}$ and $\omega_{ij} = 0$. We define the transformed four-momentum and four-position vectors as $p^{\mu}$ and $x^{\mu}$ respectively. Under this boost the four-momentum components are 

\begin{equation}
 p^{\mu} = (\gamma_{\beta}m, \gamma_{\beta}m\vec{\beta}) \hspace{2mm} \text{with} \hspace{2mm} \gamma_{\beta} = \frac{1}{\sqrt{1-\beta^{2}}}.
\label{momu}
\end{equation}    

Obviously the action on the four-momentum vector in terms of the parameter $\vec{\beta}$ is given by the standard expression. In terms of the four-momentum $p^{\mu}$, and the original vectors $q^{\mu}$ and $y^\mu$, the expression for the four-position transformation is given by

\begin{equation}
 x^{\mu}f_{\mu}{^{\nu}}(p) = \Lambda^{\nu}{_{\mu}}(\omega)y^{\alpha}f_{\alpha}{^{\mu}}(q).
\label{xtransform}
\end{equation}

This algebraic equation for the components of $x^{\mu}$ can be easily solved using the method of determinants. For this computation the particular parameterization of the function $\rho^{\mu}(p)$ given in Eq.~($\ref{RHOP}$) is crucial in order to obtain compact expressions. We obtain  

\begin{equation}
  x^{0} = \gamma_{\beta}y^{0}\frac{F'(E_{q})}{F'(E_{p})}, \hspace{2mm} x^{i} = \gamma_{\beta}y^{0}\beta^{i}\frac{F'(E_{q})}{F'(p)}.
\label{xtransformation}
\end{equation} 

Given the symmetry of our setup we obtain a modified but smooth dependence of $x^{\mu}$ on the parameters $\beta^{i}$. It is worth noting that the functions $F'$ also depend on the parameters $T$ and $L$. This dependence introduces a non-trivial connection between the position transformations of this version of DSR, and the parameters $T$ and $L$, which are directly related to the minimum uncertainties in time and position.  

In order to interpret the effects of these modified position transformations we proceed to compute the velocity associated to this particle~\cite{Hossenfelder:2006rr}. From the boosted reference frame the position and time coordinates of the particle are given by $\vec{x}$ and $x^{0}$, respectively. Considering that at the time $x^{0}=0$ the position is $\vec{x}=0$, the velocity of the particle in any future moment is defined by

\begin{equation}
 \vec{v} = \frac{\vec{x}}{x^{0}} \hspace{2mm} = \frac{F'(E_{p})}{F'(p)}\vec{\beta} \hspace{4mm}  \Rightarrow  
 \hspace{2mm} \vec{v} = \frac{F'(E_{p})}{F'(p)}\frac{\vec{p}}{E_{p}}, 
\label{vbeta}
\end{equation}

\noindent where in the second formula we express the velocity in terms of pure physical variables. This expression reduces to the standard formula for $p^{\mu}\rightarrow 0$, given that $F'\rightarrow 1$ in this limit.

For the derivation of the expression~(\ref{vbeta}) we have used a simple case, which corresponds to the transformation of a particle at rest located at the origin of coordinates. Now we would like to show that this relation is consistent for more generic situations. Consider a particle with non-trivial velocity, whose four-momentum and four-position vectors are 

\begin{equation}
  q^{\mu} = (\gamma_{\alpha}m,\gamma_{\alpha}m\vec{\alpha}) \hspace{2mm} \text{and} \hspace{2mm} y^{\mu} = (t,\vec{u}t), 
\hspace{2mm} \text{with} \hspace{2mm} \vec{u} = \frac{F'(E_{q})}{F'(q)}\vec{\alpha},
\label{withvel}
\end{equation}

\noindent where the four-momentum and four-position vectors are obtained using the previous results. Now we apply a DSR boost with parameters $\omega_{0i} = \beta_{i}$ and $\omega_{ij} = 0$ in order to observe the modifications of both the momentum and velocity with respect to this parameter. After some algebra we obtain 

\begin{eqnarray}
 p^{\mu} &=& (\gamma_{\sigma}m, \gamma_{\sigma}m\vec{\sigma}) \hspace{2mm} \text{with} \hspace{2mm} \vec{\sigma} = \frac{\vec{\alpha} + \vec{\beta}}{1+\vec{\alpha}\cdot\vec{\beta}} \hspace{2mm} 
 \nonumber \\
  \vec{x} &=& \frac{\gamma_{\beta}y^{0}F'(E_{q})(\vec{\alpha} + \vec{\beta})}{F'(p)}, \hspace{2mm} 
  x^{0} = \frac{\gamma_{\beta}y^{0}F'(E_{q})(1+\vec{\alpha}\cdot\vec{\beta})}{F'(E_{p})} \nonumber \\
 \vec{v} &=&  \frac{\vec{x}}{x^{0}} \hspace{2mm} = \frac{F'(E_{p})}{F'(p)}\vec{\sigma} \hspace{2mm} \Rightarrow \hspace{2mm} \vec{v} = \frac{F'(E_{p})}{F'(p)}\frac{\vec{p}}{E_{p}}.
\label{mom}
\end{eqnarray}     

From this result we can observe that the relation between velocity, momentum and energy is retained. Furthermore, it is possible to show that under pure rotations both the three-momentum and the three-position transform as standard vectors. Thus, we have defined completely the active transformations for this GUP/DSR approach. Regarding the interpretation of passive transformations a similar analogy between DSR and SR is not straightforward, given the nonlinear mixing between four-momentum and four-position vectors. We refer to~\cite{Hossenfelder:2006rr} and~\cite{Kimberly:2003hp} for discussions that could be applicable to this particular DSR scenario.

\section{Tackling two DSR problems}

The modified expression for the velocity gives us the option to comment on a couple of problematic issues that arise in the context of DSR transformations. The first issue is related to the consistent definition of the relative velocity between reference frames. The second is related to the transformation properties of 
macroscopic objects under DSR boosts. In order to illustrate this discussion we consider the representative function $F$ given by the first bounded function of Table~\ref{maps}. Concerning the possible values of the parameters $T$ and $L$ we start by considering the standard option, which assigns to these parameters constant values greater than $1/M_{Pl}$. Thus, the formula for the velocity is 

\begin{eqnarray}
  \vec{v} &=& \bigg{(}\frac{Lp}{TE_{p}}\bigg{)}^2\bigg{(}\frac{1-(1+2(TE_{p})^2)^{-1/2}}{1-(1+2(Lp)^2)^{-1/2}}\bigg{)}\frac{\vec{p}}{E_{p}}.
\label{absolutev}
\end{eqnarray}

From this particular example we can notice a general issue. The velocity obtained in this GUP/DSR approach in general will not depend only on the combination $(p/E_{p})$. Instead, we have an explicit dependence on $E_{p}$ and $p$, which introduces a dependence on the mass of particles. Considering this dependence on the mass we notice two problems: 

\begin{itemize}
 \item Particles that have different masses and the same velocity in a given reference frame A will be observed as moving with different velocities from another reference frame B, obtained from A through a DSR boost. This behavior makes it impossible to the observers from A and B to agree on the definition of only one relative velocity between both reference frames.
 \item Macroscopic objects, with masses in general greater than $M_{Pl}$, will have strongly modified relations between velocity and momentum. This is obviously a problem because we definitely know that heavily massive objects obey standard relativistic relations in a wide range of velocities. 
\end{itemize}

For an extended discussion of similar problems in the context of DSR, but considering modified dispersion relations in momentum space, see~\cite{AmelinoCamelia:2000mn,AmelinoCamelia:2000ge,Judes:2002bw}. In this work we consider a simple approach to account for these two issues. We have seen that the previous problematic aspects are related to the explicit velocity-mass dependence. Hence, a simple approach to circumvent these issues is eliminating this dependence, and for this task we can use the freedom that we have in the choice of the parameters $T$ and $L$. We notice that a simple definition of these parameters eliminating the dependence on the mass is

\begin{equation}
  T = \frac{1}{\alpha_{T}m} \hspace{4mm} \text{and} \hspace{4mm} L = \frac{1}{\alpha_{L}m}, 
\label{alphas}
\end{equation} 

\noindent where $m$ is the mass of the particle whose four-momentum and four-position vectors are subject to DSR transformations. The parameters $\alpha_{T}$ and $\alpha_{L}$ are dimensionless quantities. Using this procedure, the consistency of the DSR transformations affects directly the ML parameters. Furthermore, the effect on these parameters is far from trivial, because now the minimum uncertainty both in space and in time for a given particle depends on its mass.

After eliminating the velocity-mass dependence, we find that even if the expression for the velocity is deformed, this deformation is the same for particles with different masses. As particles with the same velocity in a given reference frame can be associated with the same boost parameter, the velocities observed in a different reference frame will also be the same. This mass-independent definition of the velocity allows us to define uniquely the relative velocity between two reference frames. For instance, this velocity can be computed from the transformation of the four-position vector of any particle which is at rest in one of the reference frames. Using the parameterization~(\ref{alphas}) this relative velocity is

\begin{equation}
  \vec{v}_{rel} = \bigg{(}\frac{\alpha_{T}}{\gamma_{\beta}}\bigg{)}^2\bigg{(}\frac{\gamma_{\beta}\beta}{\alpha_{L}}\bigg{)}^2\bigg{(}\frac{1-(1+2(\gamma_{\beta}/\alpha_{T})^2)^{-1/2}}{1-(1+2(\gamma_{\beta}\beta/\alpha_{L})^2)^{-1/2}}\bigg{)}\vec{\beta}.
\end{equation}

Considering the heavy mass of macroscopic objects we can see that strong modifications of the velocity-momentum relation will be expected just for highly relativistic particles, i.e., in the regime $E/m \gg \alpha_{T,L}$. Thus, we can protect the standard behavior of macroscopic particles at lower velocities, dictated by SR, just taking the parameters $\alpha_{T,L}$ reasonably large. The value of these parameters, contrary to the definition of $T$ and $L$, should be considered as universal for massive particles. For example, values of $\alpha_{T,L} \sim 10^{10}$ would produce appreciable modifications of the relativistic behavior of electrons only for $E > 10^6$ GeV, a result that agrees with the value predicted in~\cite{Li:2011ue} from a phenomenological analysis of superluminal neutrinos. Moreover, the value obtained for $\Delta x_{min}$ for electrons is consistent with limits from modified energy levels for the Hydrogen atom in the GUP scenario~\cite{Kempf:1994su,Hossenfelder:2004gj,Hossenfelder:2003jz}.

\section{ML phenomenology: superluminal neutrinos}

For the phenomenology discussion we are going to consider the definition~(\ref{alphas}) of the parameters $T$ and $L$ in terms of the quantities $\alpha_{T}$ and $\alpha_{L}$, respectively, and the expression~(\ref{absolutev}) for the velocity of a particle in terms of the energy-momentum variables. For the reach of this discussion it is enough to consider the absolute value of the velocity, which in terms of $\alpha_{T,L}$ is 

\begin{equation}
  v = \bigg{(}\frac{\alpha_{T}}{E_{p}/m}\bigg{)}^2\bigg{(}\frac{p/m}{\alpha_{L}}\bigg{)}^2\bigg{(}\frac{1-(1+2(E_{p}/m)^2 (1/\alpha_{T})^2)^{-1/2}}{1-(1+2(p/m)^2(1/\alpha_{L})^2)^{-1/2}}\bigg{)}\frac{p}{E_{p}}.
\label{vpheno}
\end{equation}

As advanced in the previous chapter, the possible values of the parameters $\alpha_{T,L}$ define the size of the relativistic ratios $E/m$ or $p/m$ that a particle needs in order to receive modifications for the velocity. From now on we are going to consider only the ratio $E/m$ because for fundamental particles $E \sim p$ in a wide range of energies. The interplay between these ratios and $\alpha_{T,L}$ are shown clearly in the Eq.~(\ref{vpheno}). Using $\alpha_{T} \sim \alpha_{L}$ we distinguish two extreme regimes:

\begin{eqnarray}
   v &\sim& \bigg{(}\frac{p}{E_{p}}\bigg{)} \hspace{42mm} \text{for} \hspace{2mm} \bigg{(}\frac{E_{p}}{m}\bigg{)}\frac{1}{\alpha_{T}} \ll 1 \nonumber \\
   v &\sim& \bigg{(}\frac{\alpha_{T}}{\alpha_{L}}\bigg{)}^2\bigg{(} 1 + \frac{m}{p}(\alpha_{L}-\alpha_{T})\bigg{)} \hspace{4mm} \text{for} \hspace{2mm} \bigg{(}\frac{E_{p}}{m}\bigg{)}\frac{1}{\alpha_{T}} \gg 1.
\label{limits}
\end{eqnarray}

The first relation corresponds to the standard relativistic behavior in the low energy limit. From the second relation we notice three possible behaviors of the velocity at high energies ($E_{p} \gg m$) depending on the hierarchy between the $\alpha_{T,L}$ parameters. 

\begin{itemize}
  \item $\alpha_{T} < \alpha_{L}$ :  strictly subluminal particles, $v < 1$.
  \item $\alpha_{T} = \alpha_{L}$ :  standard photons, $v=1$ for $m \rightarrow 0$.
  \item $\alpha_{T} > \alpha_{L}$ :  potentially superluminal particles, $v > 1$.
\end{itemize}

This correlation between the ML parameters and the behavior of the velocity is inherited from the transformation rule for the position given by Eq.~(\ref{xtransformation}). However, at this stage we can interpret more clearly the connection. The cases given above indicate that the ratio between the minimum uncertainties defines the behavior of the velocity for ultra-relativistic massive particles, and vice-versa.

Considering the strict limits on the modifications of the speed of light~\cite{Aloisio:2005rc,arXiv:1102.2784,Abraham:2008ru,Maccione:2009ju}, we are going to assume that photon related parameters are given by the second option. For massive particles we could use the conservative first option, which forces the velocity to be less than one. However, considering the recent results from OPERA~\cite{:2011zb} about neutrino superluminal velocities we are going to consider the third option for massive particles. 

In order to find the explicit values of $\alpha_{T,L}$ for the neutrino case we need to know the characteristic value of $(E/m)_{c}$ that divides the subluminal and superluminal neutrino behaviors. Using the values given in~\cite{Bi:2011nd}, which provides a summary of the experimental results from SN1987A~\cite{Longo:1987}, MINOS~\cite{Adamson:2007zzb} and OPERA~\cite{:2011zb}, we can construct the Table~\ref{benchmark}. From the values on this table we find that $10^9 \lesssim (E/m)_{c} \lesssim 10^{10}$. Thus, we have to investigate the range $10^9 \lesssim \alpha_{T,L} \lesssim 10^{10}$.

\begin{table}[ht]
        \centering
        \begin{tabular}{c|c|c|c}
           Experiment &  $<$Energy$>$ (GeV) & Relative speed ($v_{\nu}-1$) & $E/m$  \\ 
          \hline
            SN1987A & $10^{-2}$ & $< 10^{-9}$ & $10^8$\\
            MINOS   & $3$ & $(5.1 \pm 2.9) \times 10^{-5}$ & $10^{10}$\\
            OPERA   & $14$ & $(2.16 \pm 0.76 \pm 0.36) \times 10^{-5}$ & $10^{11}$\\
                    & $17$ & $(2.48 \pm 0.28 \pm 0.30) \times 10^{-5}$ & $10^{11}$\\
                    & $43$ & $(2.74 \pm 0.74 \pm 0.30) \times 10^{-5}$ & $10^{11}$
        \end{tabular}
\caption{Neutrino velocity measurements from three experiments. In the computation of $E/m$ we have used $m_{\nu} = 0.1$ eV.}
\label{benchmark}
\end{table}

We can see from Eq.~(\ref{limits}) that a potential superluminal velocity will be determined by the value of the ratio $\alpha_{T}/\alpha_{L}$. Thus, in order to account for the OPERA measurement of the neutrino velocity we use $\alpha_{T}/\alpha_{L} \sim 1 + 10^{-5}$. This ratio indicates the little degree of asymmetry in the GUP/DSR parameters that we need in order to produce the superluminal behavior. 

\begin{figure}[h]
          \centering
          \includegraphics[scale=1.6]{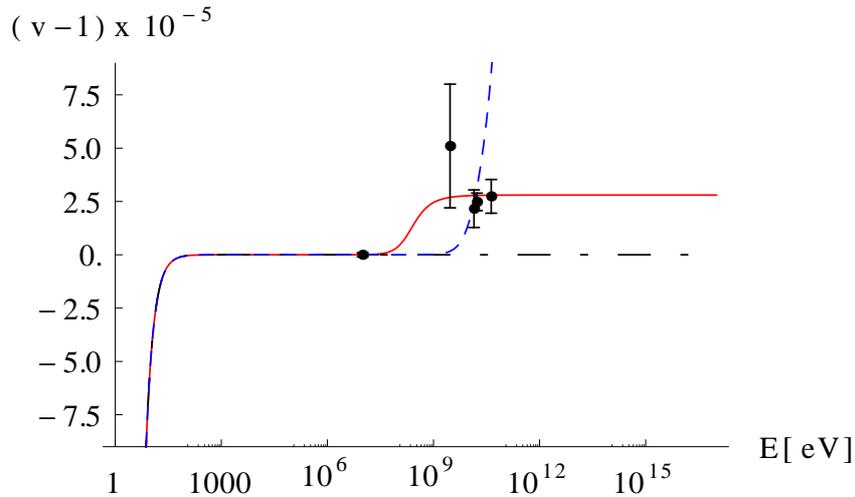}
\caption{Velocity as a function of energy for a neutrino with a mass of $0.1$ eV. The black points correspond to the experimental data from Table~\ref{benchmark}. The dashed-dot black line is the prediction from SR. The dashed blue line is the prediction of GUP/DSR using the second bounded function of Table~\ref{maps}, with $\alpha_{T}=10^{11}$ and $\alpha_{T}/\alpha_{L} = 1 + 10^{-5}$. Finally, the solid red line is the prediction of GUP/DSR using the first bounded function of Table~\ref{maps}, with $\alpha_{T}=3\times10^{9}$ and $\alpha_{T}/\alpha_{L} = 1 + 1.4 \times 10^{-5}$.}
\label{neutrino-constant}
\end{figure}

In Fig.~\ref{neutrino-constant} we plot the theoretical velocity functions corresponding to both bounded functions given in Table~\ref{maps}, together with the experimental values obtained from Table~\ref{benchmark}. Using the expression~(\ref{vpheno}), which is derived using the first bounded function of Table~\ref{maps}, we obtain a constant behavior of the velocity for very high energies, which fits the data from the experiments reasonably well. The bounded function used in this case was introduced in the context of GUP in~\cite{Kempf:1996nk}. That work does not consider superluminal velocities for ultra-relativistic neutrinos, therefore this good agreement is a coincidence. In the GUP literature we can find other bounded functions, like the second bounded function of Table~\ref{maps}, that are equivalently useful in order to find finite minimum uncertainties, however in general they are not able to reproduce the same behavior for the velocity function, as shown in Fig.~\ref{neutrino-constant}. In this sense the connection between ML and constant superluminal velocities is not one to one. A general feature in this GUP/DSR framework is the ability to accommodate superluminal particles in a reference frame invariant approach. 

In the case where the parameters $\alpha_{T,L}$ are considered as universal for every particle, independent of its mass $m_{p}$, we obtain that the limiting velocity is common, see Eq.~(\ref{limits}), but the minimum uncertainties in position and time are mass dependent, see Eq.~(\ref{alphas}). We can extract the corresponding energy at which the superluminal behaviors for particles would appear just considering that the ratio between $E/m_{p}$ and $\alpha_{T}$ at the threshold energy $E_{th}$ should be of the same order. Thus, we have that $E_{th} = m_{p} \alpha_{T}$. We obtain that for electrons the energy threshold would be $10^{6}$ GeV, for muons $10^9$ GeV and for protons $10^{11}$ GeV. These three energy values currently cannot be achieved in terrestrial experiments, however they look very accessible in cosmic ray experiments. Then in principle we have the resources to search for the confirmation of this universal behavior, or in general to falsify this particular GUP/DSR framework. 

\section{Conclusions} 

As a general consequence of a Generalized Uncertainty Principle (GUP) plus Doubly Special Relativity (DSR) framework for Minimum Length (ML) we have found interesting connections between quantum uncertainties of space-time operators and the relativistic expression for velocity in terms of energy and momentum. We notice that both the minimum lengths associated to space-time uncertainties and the velocity behavior of ultra-relativistic particles are governed by the same free parameters of the GUP/DSR framework. 

These quantum-relativistic relations allow us to accommodate the superluminal velocity of neutrinos reported by OPERA~\cite{:2011zb}, if we consider asymmetric minimum lengths in space and time. The values for the GUP/DSR parameters are computed in order to satisfy both the subluminal behavior of neutrinos in the regime $E/m \lesssim 10^8$ and the superluminal behavior for $E/m \gtrsim 10^{10}$. Furthermore, we are able to accommodate a constant velocity function in the ultra-high energy regime, which allows us to describe the data from OPERA quite well.

Considering the values of the GUP/DSR parameters obtained from the neutrino analysis as common for every particle, we are able to predict a superluminal behavior not only for neutrinos, but also for all other massive particles. In this case we obtain that the maximum velocity for different massive particles should be the same, but the energy at which the modified velocity effects would be important is proportional to the respective particle mass.

\section*{Acknowledgments}

I would like to thank J\"orn Kersten, Jan Heisig and Robert Richter for very helpful discussions. I also thanks CINVESTAV in Mexico City for hospitality during stages of this work. This work was supported by the German Science Foundation (DFG) via the Junior Research Group ``SUSY Phenomenology'' within the Collaborative Research Center 676 ``Particles, Strings and the Early Universe''.

\end{document}